\documentstyle[aps,prl,epsf]{revtex}
\begin{document}
\title{Rising Level of Public Exposure  to Mobile Phones:
Accumulation through Additivity and Reflectivity}
\author{
 Tsuyoshi Hondou}
\address{ 
         Department of Physics, Tohoku University, Sendai 980-8578,  Japan \\ 
         Institut Curie, Section de Recherche, UMR 168, 26 rue d'Ulm,
         75248 Paris Cedex 05, France
}
\maketitle      
\begin{abstract}
   A dramatic development occurring in our daily life is the increasing 
use of mobile
equipment including mobile phones and wireless access to the Internet.
They enable us to access several 
types of information more easily than in the past. 
Simultaneously, the density of mobile users is rapidly increasing.
When  hundreds of mobile phones emit radiation, their total power is found
to be comparable to that of a microwave oven
or a satellite broadcasting station.
Thus, the question arises:
what is the public exposure level in an area with many sources of
electromagnetic wave emission? We show that
this level can reach the reference level for general public exposure 
(ICNIRP Guideline) in 
daily life.
This is caused by the fundamental properties of electromagnetic field,
namely, reflection and additivity.
The level of exposure
is found to be  much higher than that estimated by
   the conventional framework of analysis that
assumes 
that the level rapidly decreases with the inverse  square
 distance between
the source and the affected person.  
A simple formula for the  exposure level is derived  by
applying  energetics to the 
electromagnetic field. The formula reveals a potential
   risk of intensive exposure.
\end{abstract}


Japan is a country where many people use mobile
communicators  frequently almost anywhere including in
public transport, where access is enabled even in underground trains
by  base stations placed in tunnels, although their use is prohibited 
in hospitals
   and airplanes to prevent  possible
fatal accidents. On one occasion the author experienced
 interference from a mobile phone in the headset of a music player, and
later found that serious interference
also occurred in some hearing aids; such interference occurs even if
 there is a
 distance of more than five meters between the source
and an affected person.
The incidents were not consistent with  a well-known paradigm for
interference by mobile  phones that emphasizes that the interference 
occurs only if the affected person is sufficiently close to  the source.
Based on this paradigm, several guidelines have been
constructed, one of which recommends, ``people should not use mobile
phones within 22
centimeters of a person with a cardiac pacemaker.''\cite{JAPAN}
This {\it short-range interference paradigm} has been widely accepted
   as a  guidelines in several places and, to our knowledge, it is the
only indication to citizens
of the possibility of intensive exposure. Extensive studies
are also in progress to clarify the effects on the health of mobile
 users themselves, who
are  the nearest to the mobiles\cite{LINKS,UK,WHO}.
Here, we do not discuss this aspect because the users use the mobiles
at their own risk. Rather, we focus  on the issue of  public exposure
to emission from anonymous users, because an affected person cannot
control or avoid  exposure even when a health condition 
exists. In this sense, the latter is a more serious problem than the former.

The lack of  recognition of the possibility of 
 high levels of exposure could be attributed to
the lack of a simple theory
describing exposure caused  by a distributed
emission of electromagnetic waves in a reflective boundary;
 the conventional estimation of exposure is based only on the
{\it short-range interaction paradigm} and is not sufficient
to find the level of exposure caused by distributed emissions.
Conventional analysis is valid
as long as there is {\it only one} 
mobile phone and also the boundary is {\it not reflective}.
These conditions are not appropriate for considering
 current situations.
Thus, we will derive a simple and generic formula
for electromagnetic field
and demonstrate the possibility of high levels of exposure
in certain situations.
Although we initiated this analysis as a result of being 
motivated by an incident in a public train,
the following formula is applicable to more general situations.
The physical quantity that we will estimate is the average equivalent
poynting vector
$P$  (W/m$^2$) which represents the energy flux,
and it is used to describe  the
   level of exposure.
We work within the framework of classical  electromagnetism\cite{phillips}.
Comparison of the result is made with international
guidelines for exposure limits, for which  
extensive studies on possible problems have been performed\cite{ICNIRP}.

Consider a closed box (system) with a reflective surface boundary,
in which sources of electromagnetic waves are spatially 
distributed.
Some parts of the boundary may be nonreflective or open.
We assume for simplicity that waves coming from different sources are incoherent.
Conservation  of energy in the electromagnetic field
leads directly to the balance equation for the electromagnetic energy, $U$, accumulated
in the system, 
\begin{equation}
\frac{d}{dt} U + J_E = W_r ,
\end{equation}
where $J_E$ and $W_r$ are an outward energy flux at the boundary and the sum
of emission powers in the system, respectively.
Our concern is only the average energy density that corresponds to
the equivalent poynting vector,  because the deviation from the average
value should be small in the case of a high reflection probability at
the boundary.
Thus, according to the conventional method of
statistical physics, we hereafter ignore the details of the geometry of the
system and
   introduce the characteristic length of the system $L^{\ast}$ that is  a
mean free path between successive reflections at the boundary,
where the electromagnetic wave is assumed to behave as a {\it ray}
 which propagates
   freely. The characteristic
time, $t^{\ast}$, is also defined through $L^{\ast}$ and the velocity
of the light, $c$, as $t^{\ast} = L^{\ast}/c $,
that is, the mean free time between reflections.
If there were no reflection at the boundary,
all of the energy in the system would diffuse 
through the boundary in time $t^{\ast}$.
Hence,  we have the relation,
\begin{equation}
t^{\ast} \cdot J_E = U \, .
\end{equation}
The effect of reflection is incorporated in the right-hand
side of eq. (1) as an additional emission term.
The power introduced into the system
by the reflection is a product of the flux
$J_E$ and the {\it average} reflection
probability\cite{average}, $R$.  With eqs. (1) and (2), we reach the balance equation
including the effect of
   reflection:
\begin{equation}
   \frac{d}{dt} U + \frac{c U}{L^{\ast}} = W_r +  \frac{c U}{L^{*}} R \, .
\end{equation}

As a stationary solution, which is realized for a timescale longer than
$t^{\ast}/k_d$, we have
\begin{equation}
U= \frac{L^{\ast} W_r}{c \, k_{d} } \, ,
\end{equation}
where $k_{d}$ is the average dissipation probability at the boundary
defined as $k_{d} \equiv 1-R$;
   its value lies between $0$ and $1$.
Summing up the possible multiple reflection terms gives the same stationary
result.
The average equivalent poynting vector, $\langle P \rangle $, is therefore
obtained through the relation
$\langle P \rangle = u c = \frac{U}{V} c$, where $u$ is the energy density
and  $V$ is the volume of the system, as
\begin{equation}
   \langle P \rangle = \frac{L^{\ast}  W_r }{V k_d} \,  .
\end{equation}
This is the main result of the study.
This simple formula predicts the average strength of the electromagnetic
 field in 
the system.
We note that the analytical expression of eq. (5) itself was derived in a 
generic situation. The reader may apply it to several specific  systems;
for example, buses, elevators, prefabricated houses, concert halls and so on.
The differences between 
the situations may be accounted for by changes in the parameters.
   We also note that the formula may be directly derived by
dimensional analysis for the system characterized by four parameters:
the mean free path $L^{\ast}$, the total emission power $W_r$,  the system
volume $V$, and the average dissipation probability $k_d$. If we introduce
the average emission power per mobile, $w_r$, the total emission power,
$W_r$,  is written as $W_r = N  w_r$, where $N$ is the number of
sources (mobiles).
\begin{figure}
\epsfile{file=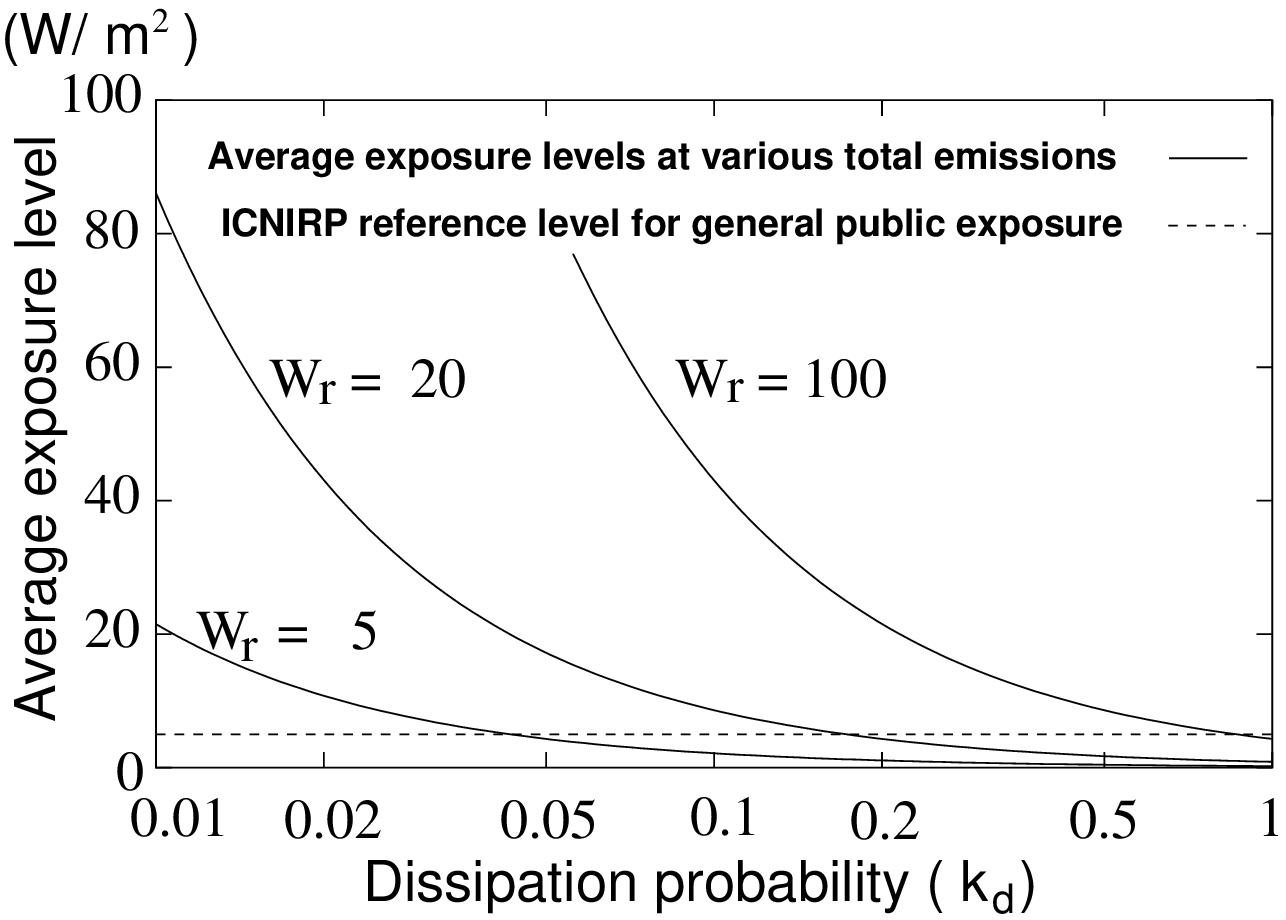,scale=0.5}
\caption{
  Average exposure level $\langle P \rangle$
is shown as a function of the average dissipation probability,
$k_d$: $\langle P \rangle = \frac{L^{\ast} W_r}{V k_d}$,
where $L^{\ast}$, $V$, $k_d$,  and $W_r$ are the
characteristic length of the system,
the volume, the average dissipation probability, and the total power of
emissions in the system, respectively.
The curves represent three different powers of total emission,
$W_r = 5, \, 20,  \mbox{ and }  100$ watts, where $V=112$
(m$^{3}$), and  $L^{\ast} = \sqrt[3]{V}
= \sqrt[3]{112} $ (m) is assumed.
   The horizontal dotted line indicates the reference level for general
public exposure
   (1 GHz) set by ICNIRP.
The reader may observe that the exposure level has dependence on
several
parameters. By scaling, one can obtain the dependence on other parameters.
}
\end{figure}

   From this result, we learn that the exposure level rapidly rises
as the dissipation probability, $k_d$, decreases to zero,
which would be the case if all boundaries were made of a metal with a reflection
probability near unity;
  an example may be an elevator without windows.
This mechanism of small dissipation in metal is already utilized in
waveguides for 
   transporting microwave with minimum loss.
On the contrary, the conventional estimation corresponds to
   the case that
$k_d =1$ in eq. (5), where the exposure level is minimum.
The level of exposure $\langle P \rangle$ increases in proportion to the total
emitted power $W_r$, which in turn increases in
 proportion to the number of 
mobiles $N$;
this {\it mass} (additive)  {\it effect} has scarcely been discussed in the
literature\cite{LINKS}.

   Although the formula is applicable to various
situations, an order estimation for a specific example is useful 
for demonstrating 
how to apply the formula to a concrete system, and 
 this result is {\it never} negligible in certain situations.
We emphasize that our aim is not to find an individual value
of exposure in a specialized situation, but to point out the
significance  of the phenomenon. Readers may perform individual
calculations using the formula (eq. (5)) with a special set of parameters.
In order to perform the order estimation, we require the
characteristic values, $L^{\ast}$, $W_{r}$, $V$ and $k_d$.
Here we confine
ourselves to the typical values for a commuter train car in Japan.
We use the values for a car in a series 3000 train of Tokyu Corporation,
one of the most recently introduced commuter trains in metropolitan
areas of Japan,
for which  $V = 112$ (m$^{3}$),  and $L^{\ast} \sim \sqrt[3]{V} = \sqrt[3]{112}
 $(m)\cite{T1}.
Since the bodies of modern trains are made of metals, the reflection
probability inside the body
 may be approximated as unity because the dissipation at 
the point of reflection is 
negligible compared to other types of dissipation, for example, 
through windows\cite{RR}. 
Because
waves are assumed to disappear irreversibly
through the window, the dissipation probability, $k_{d}$, may be estimated
in the first approximation from the ratio of the area of the windows to the
gross surface area
of the car.
In our case (Tokyu series 3000) it is estimated as $k_d=0.10$\cite{TOKYU}.
The total emission power also  depends significantly on the situation.
For greater generality, we demonstrate exposure levels
at three different values, $W_r = 5,\,  20, \,  100 $ watts,
with the reference level for general public exposure of 1 GHz given
 by the International Commission on Non-Ionizing Radiation 
Protection (ICNIRP)\cite{ICNIRP}.
    The exposure level increases as the dissipation probability decreases.
If we decrease the power, $W_r$, the critical value of the dissipation
probability decreases. In the case of a large power, $W_r =100$, the
average exposure level is of the order  of the reference level
solely  because of the mass (additive) effect of the waves
 (even without the effect
of reflection).
Since we have not yet introduced any special parameters except for the 
volume of the system, one can use this graph (Fig.1) to predict
exposure levels of other systems 
 (not only trains) of similar volume. 

To clarify how the emission power, $W_r$, corresponds to the situation
on the train, we illustrate it with a specific example. In a commuter
train, passenger  capacities are as follows: designed passenger capacity
(100\% {\it jyosharitsu}) is 151 persons, and the seating capacity
 is 54 or 51 persons\cite{TOKYU}.
Since the proportion of existing passengers to the  designed passenger
 capacity
 ({\it jyosharitsu}) can exceed 200\% under the most crowded
condition,  a car can hold more than 300 persons. 
Thus, the case $W_r = 100 $ watts corresponds, for example, to the
situation in which each of
300 mobiles 
emits  a power of $0.33$ watts; the case $W_r = 20$ watts corresponds
to the situation in which 50 mobiles emit a power of 0.4 watts each, 
 and so on.
As the average dissipation probability in the Tokyu series 3000
 is $k_d = 0.10 $, the critical total power according to 
 the ICNIRP reference level is $W_r = 12$ watts.
Note that a mobile communicator emits power to some degree 
as long as it is switched on\cite{variation}.
   For more detailed consideration, here we note  two
uncertainty factors:
i) the presence of passengers which is assumed to increase
the average dissipation probability; ii) the shielding effect of 
windows,  of which the characteristic length
is comparable to or less than the wavelength 
(0.3 m at 1 GHz).
   The former  decreases the level of exposure,
while the latter increases  it.

On the basis of this observation, we conclude that
in spite of several
uncertainty factors
there is the non-negligible possibility that the simultaneous use of
a number of mobiles in an area with a reflective boundary creates a  
  level of exposure which can exceed that stipulated in the ICNIRP guidelines.
Let us consider a commuter train that unexpectedly stops between stations 
due to an incident (typical for morning trains in 
metropolitan areas of Japan). 
What will passengers  
with mobile phones, who find themselves to be late for work, 
do? 
Once a closed area is filled with electromagnetic
waves to a considerable level, the effect of their interference on
people
with electromedical instruments could be  serious,
   because they have
no way to avoid the emissions\cite{experiment}.

 In this paper, we have analyzed how the exposure level
increases depending on the situation, by reflection and 
additivity of
electromagnetic waves.
We analytically derived a generic formula with which one can predict 
the possible
exposure level. For an example of a realistic situation, 
we applied the formula to a commuter train and performed an
order estimation
of the possible exposure level.  It was shown that 
the simultaneous use of a number of  mobile communicators
in a closed area may result in a considerable level of 
exposure, as high as that determined by the ICNIRP
to be the limit for public safety\cite{ICNIRP}.
These results are easy to understand: Consider a
   dark room  with a single
light with low power, where all of the boundaries, walls, ceiling 
and floor, are
black. If we cover the black boundaries with mirrors,
it significantly increases the brightness without
incurring any additional cost for power for lighting.
Also, if we increase the number of lights,
the brightness  will increase throughout the
room. The mechanism of the increase of intensity
is completely the same for this light and the
electromagnetic waves 
of the present case,
   because both are
governed by the same fundamental properties of electromagnetism,
namely, reflectivity and additivity. Therefore, the emergence
of the intensive field is not surprising.
Increasing public exposure to electromagnetic wave emission
 is a form of environmental pollution;
in both cases, naive personal consumption will lead to 
global pollution, which eventually may hurt
humans.
  Since the increase of electromagnetic field by reflective boundaries
and the additivity of  
sources has not  been recognized yet, further detailed studies on 
various situations and
the development of appropriate regulations are
required.

We acknowledge Mr. Tamaoki at Tokyu Corporation for providing us with 
detailed
data on their commuter trains, East Japan Railway Company for providing 
a plan of their
commuter trains,  C. F\"utterer, T. J. Harata, H. Miyata, T. Tsuzuki, 
Y. Nozue, H. Yasuhara, T. Suzuki, and  M.
Sano for helpful comments,  and Y. Hayakawa, K. Sekimoto, J. Shibata, S. Nasuno,
K. Kaneko, K. Ikeda and S. Takagi for critical reading of the manuscript
and helpful comments and encouragement, and a proofreader at Myu Research for 
great help in English expression. This work was supported in part by a 
Japanese Grand-in-Aid for the Science Research Fund from the Ministry of
Education, Science and Culture (Grant No. 12740226).


\begin{thebibliography}{99}
\bibitem{average}
 Average reflection probability is calculated, in the first approximation, 
as follows: Imagine a surface boundary at which the electromagnetic wave
is reflected (or dissipates).
Assume that the surface is made of a material which is fully reflective,
except for windows through which the wave dissipates.
Assume that the ratio of the area of the windows to the total surface is 10\%. 
Then, the {\it average} reflection probability, $R$, is approximated as 0.9.
Thus, the average dissipation probability $k_d = 1 - R = 0.1$.
\bibitem{RR} The reflection probability, in principle, depends strongly 
on the 
conditions. However, in the microwave frequency region, it is known 
that the reflection probability may be treated as almost unity 
in comparison with other dissipation 
processes: Let us consider the reflection probability  of the electromagnetic wave
propagating perpendicular to the surface of the reflective
boundary.
The reflection probability, $r$,  can be estimated using the formula applicable
to the wave with length longer than that of visible light,
$r = 1 - 2 \sqrt{2 \epsilon_0 \omega / \sigma}$,
where $\epsilon_0$, $\omega$, and $\sigma$ are the dielectric constant
of vacuum, the angular frequency of an electromagnetic wave and 
the electric conductivity, respectively 
\cite{phillips}.
By using an experimental value for the electric conductivity\cite{RIKA},
we obtain the reflection probability. For aluminum, it is 0.9999.
Since it is known that the values of electric conductivity for most 
 metals lie in the order of approximately
$10^{7}$ ($\Omega^{-1} m^{-1}$), 
the reflection probability is estimated as approximately
0.9998. This shows that the dissipation of the wave 
upon reflection
 is negligible
compared to other dissipation processes in the present case.
\bibitem{variation}
 The output of a mobile communicator may depend on the intensity 
of the wave from a base station. 
As shown by the ratio of the window area to the total body surface
of the car 
which corresponds to the average dissipation probability, the cabin of 
a train is largely shielded electromagnetically. 
Thus, the wave from the base station is considerably weakened
inside the car. It is known that
the weaker the  intensity
of the wave 
that reaches the mobile phone from the base station, the more 
power the mobile phone emits.
 We also note that the maximum output power of a mobile phone is 0.6 watts 
for analog phones, and 0.8 watts for digital phones (burst value) 
in Japan\cite{JAPAN}. In the report by W.H.O., the maximum output power
is stated to  
 range
 from
0.2 watts to 0.6 watts\cite{WHO}.
\end{thebibliography}
\end{document}